\documentclass[%
 reprint,
 amsmath,amssymb,
 aps,
]{revtex4-1}

\usepackage{graphicx}
\usepackage{dcolumn}
\usepackage{bm}
\usepackage[utf8]{inputenc}
\usepackage{xcolor}
\usepackage[euler]{textgreek}
\usepackage[normalem]{ulem}
\usepackage{eqnarray,amsmath}
\usepackage[english]{babel}

\begin{document}

\preprint{APS/123-QED}

\title{Long-time relaxation dynamics in nematic and smectic liquid crystals of soft-repulsive colloidal rods}


\author{Daniela Cywiak}
\email{d.cywiak.cordova@ugto.mx}
\affiliation{División de Ciencias e Ingenierías, Universidad de Guanajuato, Campus Le\'on, M\'exico}
 
\author{Alessandro Patti}
\email{alessandro.patti@manchester.ac.uk}
\affiliation{Department of Chemical Engineering and Analytical Science, University of Manchester, United Kingdom}

\author{Alejandro Gil-Villegas}
\email{gil@fisica.ugto.mx}
\affiliation{División de Ciencias e Ingenierías, Universidad de Guanajuato, Campus Le\'on, M\'exico}

\date{\today}

\begin{abstract}
Understanding the relaxation dynamics of colloidal suspensions is crucial to identify the elements that influence the mobility of their constituents, assess their macroscopic response across the relevant time and length scales, and thus disclose the fundamentals underpinning their exploitation in formulation engineering. In this work, we specifically assess the impact of long-ranged ordering on the relaxation dynamics of suspensions of soft-repulsive rod-like particles, which are able to self-organise into nematic and smectic liquid-crystalline phases. By performing Dynamic Monte Carlo simulations, we analyse the effect of translational and orientational order on the diffusion of the rods along the relevant directions imposed by the morphology of the background phases. To provide a clear picture of the resulting dynamics, we assess their dependence on temperature, which can dramatically determine the response time of the system relaxation and the self-diffusion coefficients of the rods. The computation of the van Hove correlation functions allows us to identify the existence of rods that diffuse significantly faster than the average and whose concentration can be accurately adjusted by a suitable choice of temperature.
\end{abstract}

\maketitle



\section{Introduction \protect  }

The study of the dynamical properties of colloidal suspensions has generated an exceptional insight into multiple fields of science and technology, shedding light on processes such as self-assembly and nucleation \cite{Pine2013a}, and systems, such as active matter \cite{Huber2018} and liquid crystals (LCs) \cite{Fraden1989, Fraden2006, Lettinga2011}, whose key mechanisms are dramatically influenced by the particle dynamics over time. The relatively recent development of experimental techniques for the synthesis of a fascinating spectrum of particle shapes \cite{Pine2013b} has boosted the interest in extending to anisotropic particles methods predominantly used to study dynamic arrest and glass transition in suspensions of spherical particles \cite{Wagner2020}. These methods include the use of hard-core models to mimic the essential features of a spectrum of colloidal systems, including LCs. As Onsager demonstrated in his seminal theory, these relatively simple models are indeed able to predict the isotropic-to-nematic transition of infinitely long rod-like particles, as later on verified by computer simulation \cite{Frenkel1988, Jackson1996}. Nevertheless, Van der Waals and long-range electrostatic forces modify the phase diagram and introduce important effects in the thermodynamic and structural properties of these systems, particularly in the location of phase boundaries, such as isotropic-nematic and nematic-smectic transitions \cite{McGrother1998, Duijneveldt2000}.

A wide variety of hard- and soft-core models have been used to gain a better insight into the behavior of thermotropic (molecular) and colloidal LCs \cite{Allen2019}. In the case of spherical particles, Medina-Noyola and coworkers demonstrated that structural properties and diffusion coefficients of spheres interacting with a repulsive Sutherland potential can be mapped onto the corresponding properties of hard-sphere (HS) systems, using effective diameters dependent on density and temperature \cite{Medina2003, Medina2011}. More recently, Jackson and coworkers obtained an effective soft-repulsive potential based on the Mie model in order to reproduce structural properties of the HS system \cite{Jackson2012}, a procedure that has been extended to square-well potentials \cite{Benavides2018} and applied successfully in molecular simulations of colloidal systems \cite{CastanedaP2018}. The approach of using potential models with soft repulsive and attractive interactions of variable range, such as the Mie model for chain molecules \cite{Davies1998, Lafitte2013}, have had important implications in the prediction of a wide variety of phase diagrams for molecular fluids \cite{Dufal2015}.

When investigating the dynamics of hard-core particles, the usual simulation techniques of choice, namely Molecular dynamics (MD) and Brownian dynamics (BD), cannot be directly applied, as the form of the interaction potential does not allow one to integrate the equations of motion. Additionally, while MD can accurately reproduce the deterministic dynamics of atoms and molecules, it is not suitable to mimic the Brownian motion of colloids unless both dispersed and continuous phases are explicitly incorporated, resulting in a very computationally demanding system to study. On the other hand, BD relies on stochastic equations of motion that implicitly incorporate the presence of a solvent, but faces, along with MD, the limitation of an integration time-step that should be sufficiently small to guarantee precision at the cost of missing the long-time relaxation dynamics. This is not a major problem in sufficiently dilute colloidal suspensions, but it is indeed a challenge in dense colloids, including LCs, whose structural relaxation completely unfolds over relatively long time scales. These limitations can be bypassed by the Dynamic Monte Carlo (DMC) method, which is able to capture the Brownian dynamics of colloids without employing stochastic or deterministic equations of motion. Based on the standard Metropolis algorithm \cite{Metropolis1953}, DMC can quantitatively and qualitatively reproduce BD simulation results in the limit of small displacements \cite{sanz2010, sanz2011, Patti2012, Cuetos2015, Corbett2018, Chiappini2020, Garcia2020, garcia2021}.

In this work, we apply the DMC method to study the dynamics of nematic (N) and smectic (Sm) LCs of colloidal rod-like particles. Rods are here represented as soft repulsive spherocylindrical (SRS) particles interacting via the Kihara potential, which has been used in the past to investigate the behavior of prolate \cite{Cuetos2005, Cuetos2015_2} and oblate \cite{Cuetos2002} spherocylinders. We are specifically interested in assessing the effect of temperature on the long-time relaxation dynamics of SRS particles in N and Sm phases. While the former are characterised by a mere orientational order with the rods approximately aligned along a common direction, but randomly positioned, the latter also exhibit translational ordering with the rods arranged in contiguous layers parallel to each other. 

This paper is organised as follows. In Section II, we describe the SRS model, provide the main details of the DMC simulation method applied in this work and introduce the observables calculated to characterize the dynamics in N and Sm LCs and its dependence on long-range ordering and temperature. In Section III, we discuss the ability of SRS to diffuse in these LC phases and the structural relaxation of the systems over time. Finally, we draw our conclusions in Section IV.


\section{\label{sec:level1.1}Model and simulations\protect}

As mentioned above, colloidal rods are here modelled as SRS particles interacting via the Kihara potential. In particular, we studied systems of $N_r=1000$ prolate spherocylinders of length-to-diameter ratio $L^{\star} \equiv L/\sigma = 5$, where $L$ and $\sigma$ are, respectively, the length and diameter of a cylindrical body capped by two hemispheres of identical diameter. The total length of the rod is thus $L+\sigma$. The interaction between particles is described by a repulsive, truncated and shifted Kihara potential that reads

\begin{equation}
   U_{ij}=\left\{
  \begin{array}{ll}
    4\epsilon\left[\left(\frac{\sigma}{d_m}\right)^{12}-\left(\frac{\sigma}{d_m} \right)^{6}+\frac{1}{4}  \right]& \quad d_m \leq \sqrt[6]{2}\sigma\\\\
    0              &      \quad d_m  > \sqrt[6]{2}\sigma
\end{array}, \right.
\end{equation}

\noindent where $U_{ij}=U_{ij}(\textbf{r}_{ij}, \hat{\bf u}_i, \hat{\bf u}_j)$. The subscripts $i$ and $j$ refer to a pair of interacting spherocylinders, $\textbf{r}_{ij}$ is the centre-to-centre distance between them, $\epsilon$ their interaction strength, $\hat{\bf u}_i$ and $\hat{\bf u}_j$ indicate the particles' orientation, and $d_m=d_m(\textbf{r}_{ij}, \hat{\bf u}_i, \hat{\bf u}_j)$ is the minimum distance between $i$ and $j$. The interested reader is referred to Ref.\,\cite{VEGA199455} for additional details on the computation of the minimum distance between prolate spherocylinders. All simulations have been performed in the canonical ensemble at constant number of particles, temperature ($T$) and volume ($V$). The density, $\rho$, of N and Sm phases was set according to the phase diagrams of Kihara spherocylinders reported in Ref.\,\cite{Cuetos2015}. In particular, $\rho_{\rm N}^{\star}=0.12$ and $\rho_{\rm Sm}^{\star} =0.15$ for N and Sm phases, respectively, with $\rho^{\star}=N_r\sigma^3/V$. To equilibrate N and Sm phases at these values of density, we melted initial configurations of perfect crystals at the desired temperature by performing standard MC simulations. In particular, rotational and translational movements of randomly selected particles were accepted or rejected according to the Metropolis algorithm \cite{Metropolis1953}. The systems were considered to be at equilibrium when the potential energy of the system had reached a steady value within moderate statistical fluctuations.

After equilibration, we performed DMC simulations to study the relaxation dynamics of N and Sm phases.  
In DMC simulations, to realistically mimic the Brownian motion of colloidal particles, unphysical moves, such as swaps or cluster moves, are not performed. An insightful description of the DMC method applied in this work is available elsewhere \cite{Patti2012, Cuetos2015, Corbett2018, Garcia2020, Chiappini2020, garcia2021}. Here we only provide a brief overview of the main features of the method and refer the interested reader to these works for details.
DMC simulations were performed in N and Sm phases at  scaled temperatures $T^{\star}=5, 8, 10, 12, 15$ and 20, with $T^{\star}=k_BT/\epsilon$ and $k_B$  the Boltzmann constant. We selected $\sigma$, $\epsilon$ and $\tau = \sigma^2/D_0$ as units of length, energy and time, respectively, where $D_0=k_BT/\mu\sigma$ is a diffusion constant and $\mu$ the viscosity of the implicit solvent. One DMC cycle consists of $N_r$ attempts of simultaneously displacing and rotating a randomly selected particle. These moves are accepted according to the probability $\min[1,\exp(-\Delta U/k_BT)]$, where $\Delta U$ is the energy difference between new and old configurations. The magnitude of elementary displacements and rotations is defined according to the Einstein equations and the particle diffusion coefficients at infinite dilution. More specifically, displacements in the direction of $\hat{\bf u}_i$ and perpendicular to it where randomly selected from uniform distributions that satisfy the conditions $|X_\parallel| \leq \sqrt{2D_\parallel \delta t_{\rm MC}}$ and $|X_\perp| \leq \sqrt{2D_\perp \delta t_{\rm MC}}$, respectively, where $\delta t_{\rm MC}$ is the arbitrarily set MC time step, while $D_\parallel$ and $D_\perp$ correspond to the translational diffusion coefficients at infinite dilution along $\hat{\bf u}_i$ and perpendicularly to it, respectively. A similar approach was used to calculate particle rotations, where the vector $\hat{\bf u}_i$ changes to $\hat{\bf u}_i+\delta\hat{\bf u}_i$, with $\delta\hat{\bf u}_i=Y_{\varphi,1} \hat{\bf w}_{j,1}+Y_{\varphi,2} \hat{\bf w}_{j,2}$. The two randomly chosen vectors $\hat{\bf w}_{j,m}$ are perpendicular to each other and to $\hat{\bf u}_i$. Finally, $Y_{\varphi,1}$ and $Y_{\varphi,2}$ are random numbers selected from uniform distributions that satisfy the condition $|Y_{\varphi,m}|\leq \sqrt{2D_\varphi \delta t_{\rm MC}}$ where $D_\varphi$ is the rotational self-diffusion coefficient of the rod.

Translational and rotational diffusion coefficients of rod-like particles at infinite dilution have been estimated by applying the analytical expressions based on the induced-forces method by Bonet Avalos and coworkers \cite{Bonet1994}. In particular:

\begin{align}
    \label{eq30}
    \frac{D_{\perp}}{D_0}&=\frac{\ln(2/\gamma)-1/2-I^{tt}}{2\pi/\gamma}\\
    \label{eq31}
    \frac{D_{\parallel}}{D_0}&=\frac{\ln(2/\gamma)-3/2-I^{tt}}{\pi/\gamma}\\
    \label{eq32}
    \frac{D_{\varphi}}{D_0}&=3\frac{\ln(2/\gamma)-11/6-I^{rr}}{\pi\sigma^2/\left(2\gamma\right)^3}
\end{align}

\noindent where $1/\gamma=2\left(L^{\star}+1\right)$, $I^{tt}\equiv\frac{1}{2}\int_{-1}^{1}dx \ln h(x)$, $I^{rr}\equiv\frac{3}{2}\int_{-1}^{1}dx x^2\ln h(x)$ and $h(x)=(1-x^{2n})^{1/2n}$. The function $h(x)$, applied to model particles with revolution symmetry, approximates very well the shape of a spherocylinder when $n=8$. Under these conditions, $I^{tt}\simeq-0.0061$ and $I^{rr}\simeq-0.017$. The translational and rotational diffusion coefficients at infinite dilution used in our DMC simulations are given in Table I for different values of the reduced temperature. We note that the value of the three sets of diffusion coefficients are different from those employed in Ref.\,\cite{Patti2012}, especially so at $T^{\star}=20$. Such difference is due to the fact that Ref.\,\cite{Patti2012} made use of the analytical expressions available for prolate spheroids, whose geometry is roughly similar, but not identical to that of spherocylinders. 

\begin{table}[ht]
\caption{Translational and rotational diffusion coefficients at infinite dilution as obtained from Eqs.\,(2-4) at different values of the reduced temperature.}
\begin{ruledtabular}
\begin{tabular}{lccc}
\textrm{$T^{\star}$}&
\textrm{$D_{\parallel}/{D_0}$}&
\textrm{$D_{\perp}/{D_0}$}&
\textrm{$D_{\varphi}/{D_0}$}\\
\colrule
5 & 0.223 & 0.177 & 0.029 \\
8 & 0.357 & 0.284 & 0.047 \\
10 & 0.446 & 0.355 & 0.059\\
12 & 0.536 & 0.427 & 0.071 \\ 
15 & 0.670 & 0.533 & 0.089 \\
17 & 0.759 & 0.605 & 0.101\\
20 & 0.893 & 0.711 & 0.119\\
\end{tabular}
\end{ruledtabular}
\end{table}

In order to obtain the actual time scale for Brownian dynamics, we have rescaled the MC timescale via the acceptance rate using the following relation

\begin{equation}
t_{\rm BD}=\frac{\mathcal{A}}{3}C_{\rm MC}\delta t_{\rm MC}
\end{equation}

\noindent where $t_{\rm BD}$ is the Brownian time after $C_{\rm MC}$ cycles, $\mathcal{A}$ is the average acceptance rate over this number of cycles and $\delta t_{MC}$ the corresponding MC time. This result allows one to rescale the dynamical properties and consistently compare the structural relaxation of N and Sm phases across the spectrum of temperatures studied here. More specifically, we evaluated the mean-squared displacement (MSD), the self-part of the Van Hove correlation function (s-VHF) and the self-part of the intermediate scattering function (s-ISF). Each of these functions has been calculated in the direction of the nematic director and in planes perpendicular to it. In particular, the parallel and perpendicular MSD read

\begin{equation}
  \big \langle \Delta r^2_{\parallel}(t)\big \rangle=\frac{1}{N_r}\bigg \langle\sum [(\mathbf{r}_j(t)-\mathbf{r}_j(0))\cdot \mathbf{n}]^2\bigg \rangle
\end{equation}

 \begin{equation}
 \big \langle \Delta r^2_{\perp}(t)\big \rangle=\frac{1}{N_r}\bigg \langle\sum [(\mathbf{r}_j(t)-\mathbf{r}_j(0)])\times \mathbf{n}]^2\bigg \rangle
\end{equation}

\noindent where $\big \langle ...\big \rangle$ denotes ensemble average and $\mathbf{n}$ is the nematic director. Similarly, the pair distribution functions parallel and perpendicular to the director are given by

\begin{equation}
    g_{\parallel}(z)=\frac{1}{N_r^2}
    \bigg \langle\sum\sum
    \delta (
    |(\mathbf{r}_i-\mathbf{r}_j)\cdot\mathbf{n}|-z)
    \bigg \rangle
\end{equation}
\begin{equation}
    g_{\perp}(r)=\frac{1}{N_r^2}
    \bigg \langle\sum\sum
    \delta (
    |(\mathbf{r}_i-\mathbf{r}_j)\times
    \mathbf{n}|-r)
    \bigg \rangle
\end{equation}

\noindent where $z$ and $r \equiv \sqrt{x^2+y^2}$ are the particle center-to-center distances in the direction of $\mathbf{n}$ and perpendicularly to it, respectively, and $\delta$ is the Dirac delta function. Additionally, the probability distribution of the particles at time $t_0$, given by the self-part of the van Hove Function, reads 
 
\begin{equation}
G_s(z,t)=\frac{1}{N_r}\bigg<\sum_{j=1}^{N} \delta
(z-[{z}_j(t+t_0)-{z}_j(t_0))]\bigg>
\end{equation}

\begin{equation}
G_s(r,t)=\frac{1}{N_r}\bigg<\sum_{j=1}^{N} \delta
(r-[{r}_j(t+t_0)-{r}_j(t_0))]\bigg>
\end{equation}

\noindent Finally, the s-ISF gives a measure of the structural relaxation of the system over time and quantifies the decay of its density fluctuations. In this case, we also estimate the parallel and perpendicular contributions as follows

\begin{equation}
F_{s,z}(t)=\frac{1}{N_r}\bigg<\sum_{j=1}^{N_r} \exp[i\mathbf{q}
\cdot[\mathbf{r}_j(t+t_0)-\mathbf{r}_j(t_0))]\cdot\mathbf{n}]\bigg>
\end{equation}

\begin{equation}
F_{s,xy}(t)=\frac{1}{N_r}\bigg<\sum_{j=1}^{N_r} \exp[i\mathbf{q}
\cdot[\mathbf{r}_j(t+t_0)-\mathbf{r}_j(t_0))\times
    \mathbf{n}]\bigg>
\end{equation}

\noindent where the wave vector $\mathbf{q}=\mathbf{q}_{\parallel}+\mathbf{q}_{\perp}$
is defined at the main peaks of the static structure factor, taking values of $|\mathbf{q}_{\parallel}|\sigma\simeq1$ and  $|\mathbf{q}_{\perp}|\sigma\simeq6$ for the parallel and perpendicular components with respect to the director, respectively.

 
\section{Results \protect  }

Before presenting and discussing the relevant observations on the dynamics of SRS rods in N and Sm phases, we first analyse the structural properties that help us distinguish between positionally and merely orientationally ordered phases. To this end, we computed the pair-correlation functions $g_\perp(r)$ and $g_\parallel(z)$, shown in Figs.\, 1 and 2, respectively. Our results indicate the existence of a typical fluid-like behavior in the perpendicular direction of both N and Sm phases, with the first peak occurring at a distance of approximately one diameter length and the amplitude of  oscillations decaying exponentially to 1 at relatively short distances. Upon increasing temperature from $T^{\star}=5$ to 20, all peaks tend to flatten and fluctuations decay at progressively shorter distances. As far as the parallel pair correlation function, $g_\parallel(z)$, is concerned, the pronounced periodic correlations of Fig.\,2 confirm the layered structure typically observed in Sm LCs. In particular, the position of each peak roughly corresponds to the location of each smectic layer, where particle density is maximal. By contrast, the inter-layer spacing is almost completely empty, the probability of observing particles in between layers being very low. Increasing temperature has a significant effect on the density distribution in the Sm phase, but a practically negligible effect in the N phase, which shows no evidence of positional order at all temperatures studied.

\begin{figure}[ht]
\includegraphics[width=8.6cm]{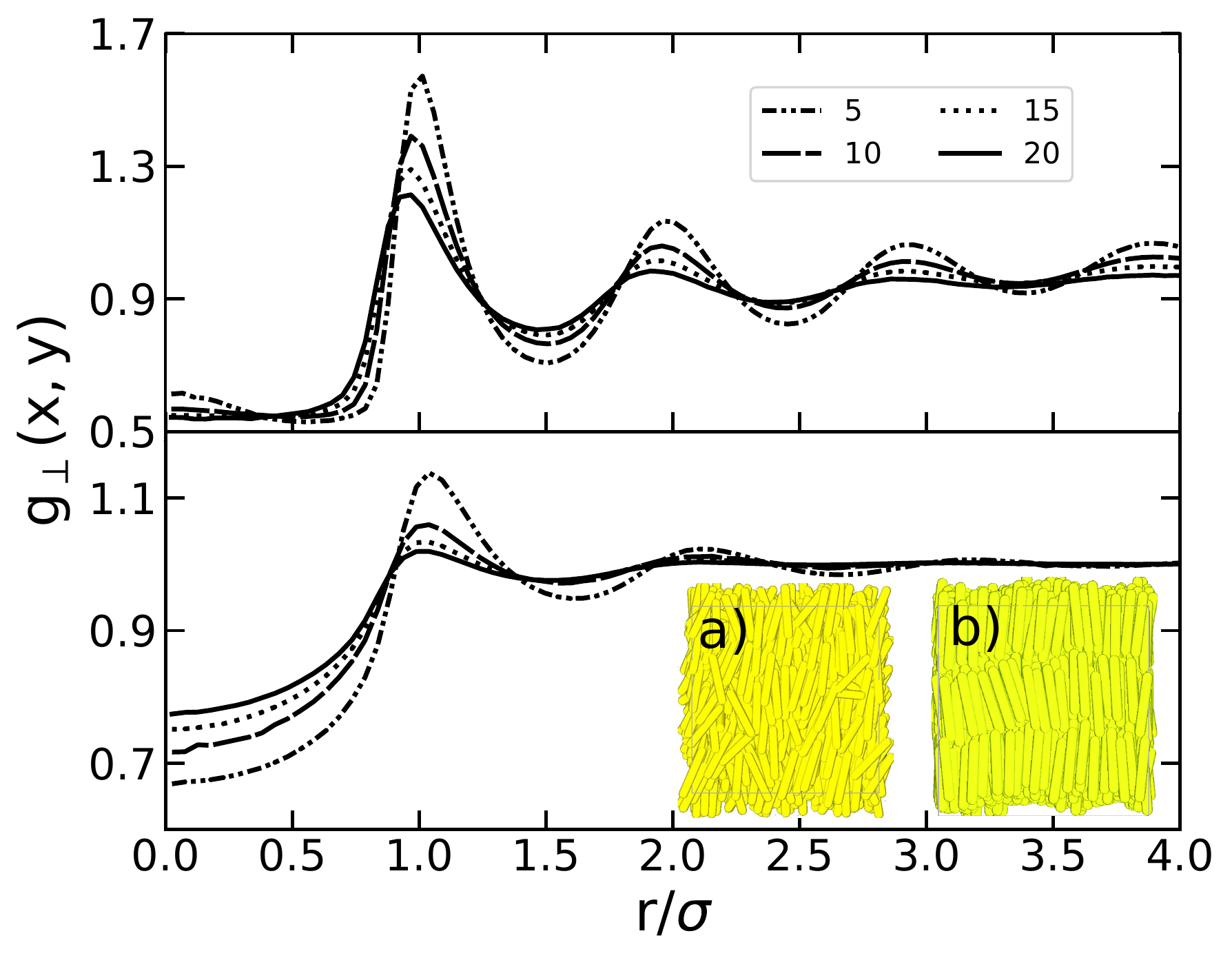}
\caption{Pair correlation functions in the direction perpendicular to the nematic director as computed in N (top) and Sm (bottom) phases at different temperatures. Insets a) and b) in the bottom frame correspond, respectively, to N ($\rho^*=0.12$) and Sm ($\rho^*=0.15$) phases at $T^{\star}=10$.}
\end{figure}

\begin{figure}[ht]
\includegraphics[width=8.6cm]{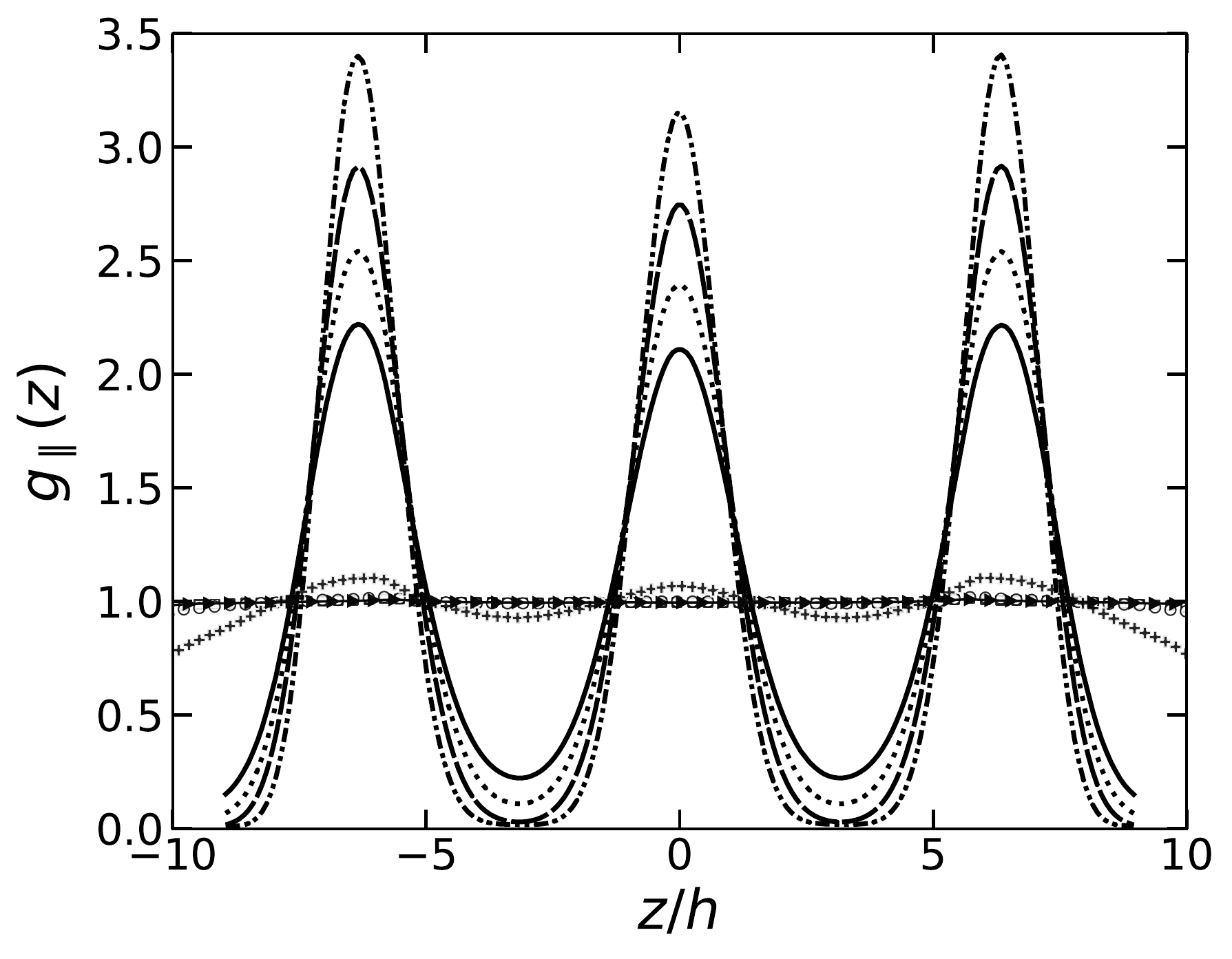}
\caption{Pair correlation functions in the direction parallel to the nematic director at $T^{\star}$=5, 10, 15 and 20. Lines and symbols refer to Sm and N phases, respectively. Plots with larger amplitudes correspond to lower temperatures. Equilibrium densities are $\rho^{\star}=0.12$ for the N phases and $\rho^{\star}=0.15$ for the Sm phases.}
\end{figure}

\begin{figure}[ht]
\includegraphics[width=8.6cm]{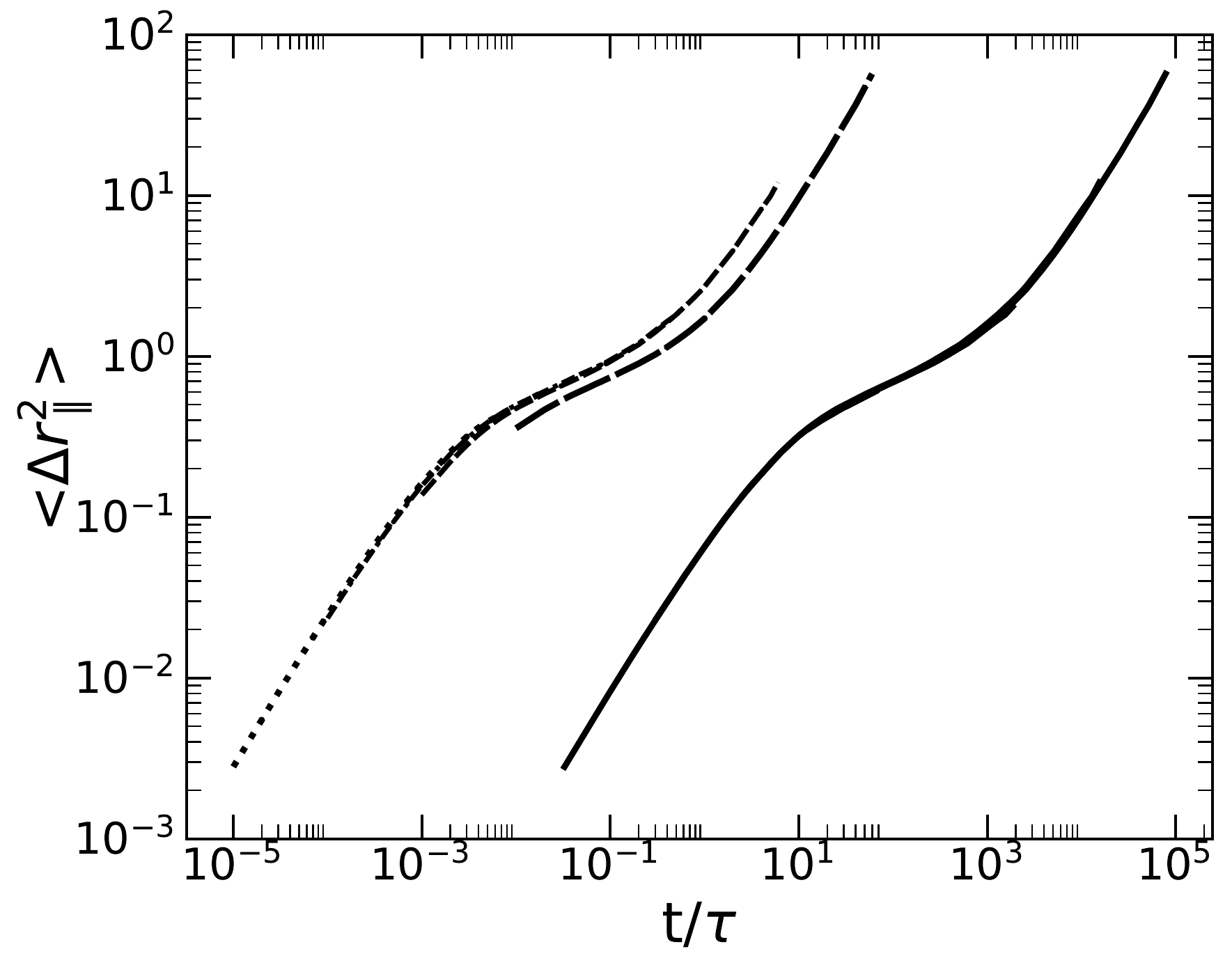}
\caption{Mean square displacement in the Sm phase at $T^{\star}=10$ in the direction parallel to the nematic vector. Dotted, dashed and long-dashed lines refer to independent DMC simulations at $\delta t_{\rm MC}/\tau=10^{-5}$, $10^{-3}$ and $10^{-2}$, respectively. Solid lines have been obtained by rescaling the corresponding dashed lines according to Eq.\,(5).}
\end{figure}

Having established the structural differences between the N and Sm phases studied here, we now investigate how these distinctive morphological attributes can in turn determine distinct dynamical signatures. To start with, we first estimated the MSD in the direction parallel and perpendicular to the nematic director and calculated the long-time diffusion coefficients. To this end, we ran DMC simulations at different values of the MC time step, between $\delta t_{\rm MC}/\tau=10^{-5}$ and $10^{-2}$, and then applied Eq.\,(5) to rescale the results and recover the unique BD time scale \cite{Patti2012}. The rescaled MSDs collapse into a single master curve as shown in Fig.\,3, where we report the parallel MSD in the Sm phase at $T^{\star}=10$. The re-scaling procedure to obtain the MSD across the whole spectrum of relevant time scales is the same for all the remaining systems. The resulting master curve has been obtained by superimposing four separate rescaled MSDs (dashed lines) calculated over DMC simulations at $\delta t_{\rm MC}/\tau=10^{-5}$, $10^{-4}$, $10^{-3}$ and $10^{-2}$. The so-calculated MSD exhibits an initial diffusive regime, mostly determined by the particle geometry, followed by an intermediate time regime where the presence of  neighboring layers, forming a sort of cage around the particles, slows down diffusion and, finally, a long-time diffusive regime that fully develops at $t/\tau>1$. Very similar tendencies have also been detected at $T^{\star}=5$, 8, 12, 15 and 20, with some differences observed in the extension of the cage effect and the onset of the long-time diffusive regime. In particular, the effect of temperature on the MSD is clarified in Fig.\,4, where we report parallel and perpendicular MSDs in N (top frame) and Sm (bottom frame) LCs at $T^{\star}=5$ and 20, that is the lowest and highest temperatures studied. For the sake of clarity, we do not show the parallel and perpendicular MSDs at intermediate temperatures, which exhibit a profile in between those reported in Fig.\,4 for N and Sm phases.
\begin{figure}[ht]
\includegraphics[width=8.6cm]{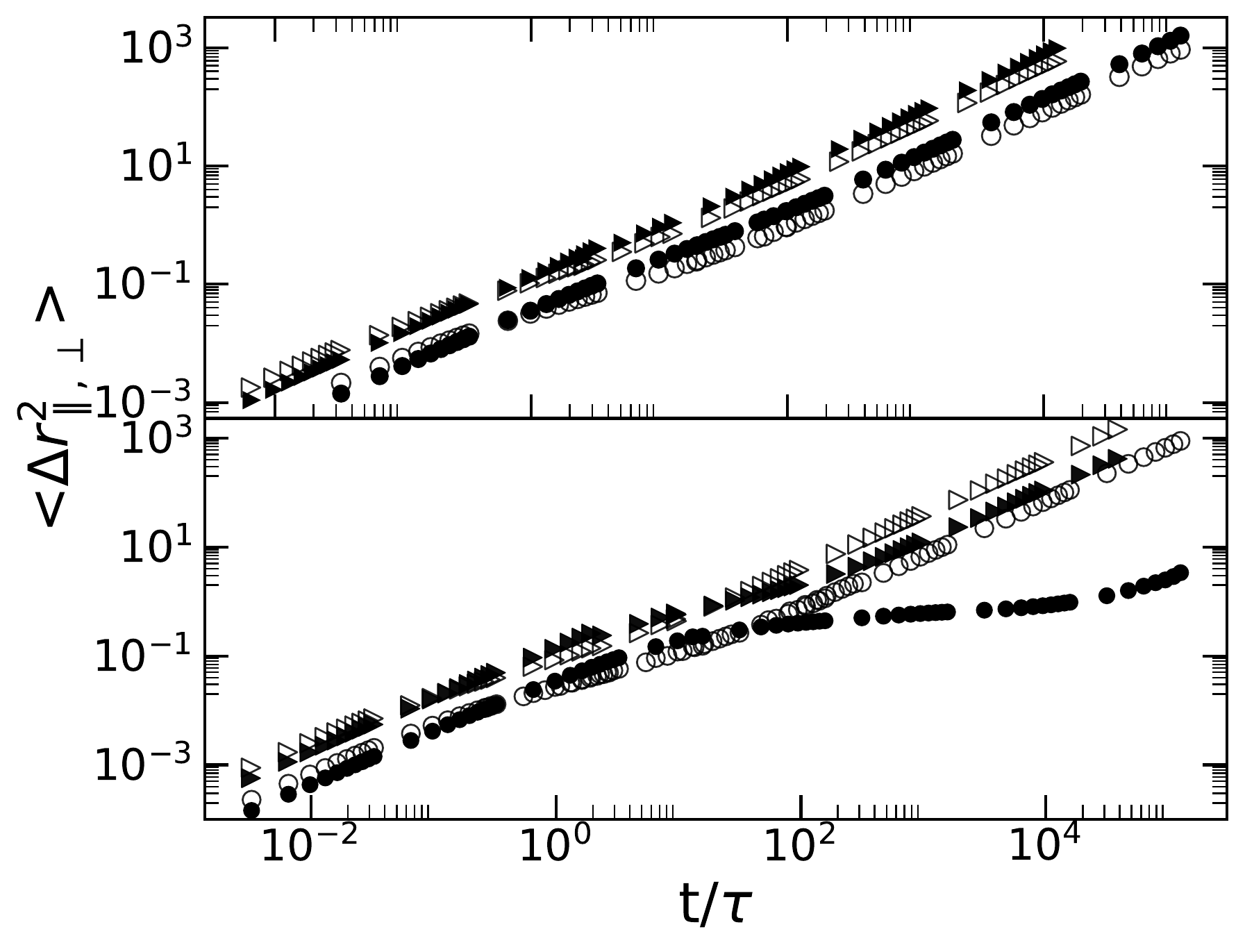}
\caption{Mean square displacements in N (top frame) and Sm (bottom frame) LCs at $T^{\star}$=5 (circles) and 20 (triangles). Open and solid symbols refer to perpendicular and parallel direction, respectively. }
\end{figure}
We also notice that, in the N phase, at short time scales the relationship between parallel and perpendicular MSDs is such that $\xi \equiv \Delta r^2_{\perp}/\Delta r^2_{\parallel}>1$, while it inverts at intermediate time scales, when the parallel MSD becomes larger and stays so up to the long-time diffusive regime. This behavior has also been reported in Brownian dynamics simulations of SRS rods at $T^{\star}=1.465$ \cite{morillo2019}, a temperature at which the phase behavior of soft spherocylinders can be mapped on that of hard spherocylinders \cite{Cuetos2015_2}. The dominant character of the long-time parallel diffusion has also been observed experimentally in N phases of rod-like viruses \cite{lettinga2005}, but less clear is whether or not this tendency already exists at short time scales as observed in simulations. To gain an insight into the effect of temperature on the ratio between perpendicular and parallel diffusivities, we have calculated the time when the crossover from $\xi>1$ to $\xi<1$ is observed. Such an inversion time, referred to as $t_i$, does indeed change with temperature and the results are shown in Fig.\,5. We observe that $t_i$ is relatively large at low temperatures and then gradually decreases following an exponential law of the type $t_i/\tau = A+B\exp(-CT^{\star})$, where $A=0.40$, $B=1.51$ and $C=0.37$ are fitting parameters. It is evident that at large enough temperatures, $t_i$ tends to a constant value, approximately equal to $0.40\tau$, that will not change significantly up to the I-to-N transition temperature.

\begin{figure}[ht]
\includegraphics[width=0.95\linewidth,height=0.25\textheight]{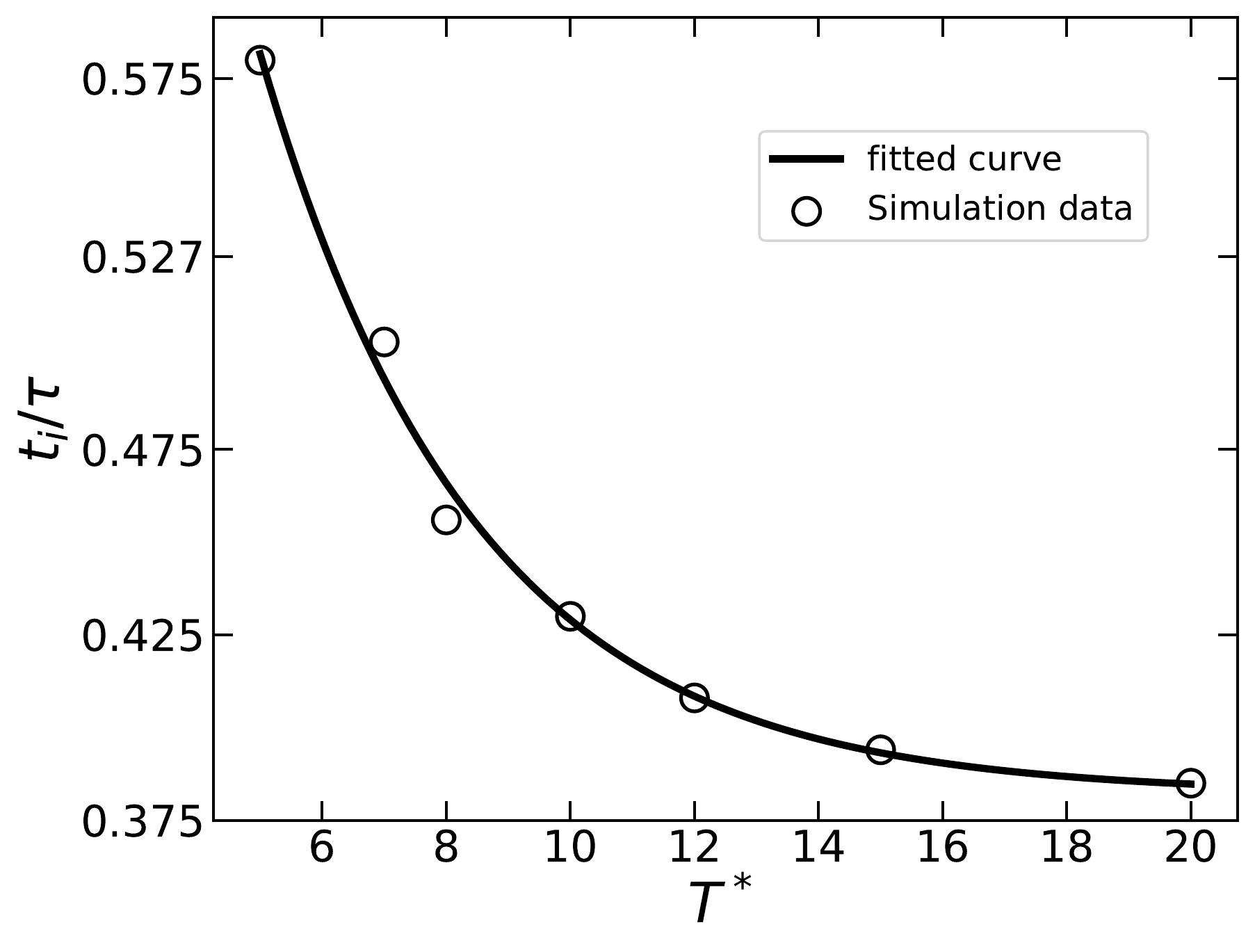}
\caption{Inversion time between parallel and perpendicular MSD of the N phase. Symbols are simulation results and the solid line is a fitting exponential function.}
\end{figure}

The MSDs are instrumental to calculate the long-time self-diffusion coefficients and their dependence on temperature. More specifically, parallel and perpendicular self-diffusion coefficients in N and Sm phases were obtained from the slope of the corresponding MSDs in the long-time diffusive regime: 

\begin{equation}
D_{\parallel, \perp}=\lim_{t \to \infty} \frac{1}{2l}\frac{d\bigg \langle\sum [\mathbf{r}_l(t)-\mathbf{r}_l(0)]^2\bigg \rangle}{dt}
\end{equation}

\noindent where $l=1$ or 2 denotes the dimensionality of particle dynamics associated to the parallel or perpendicular MSD, respectively. By contrast, the total self-diffusion coefficient has been calculated as $D_{Tot}=(D_{\parallel}+2D_{\perp})/3$. The dependence of the self-diffusion coefficients on temperature in N and Sm phases is presented in the two frames of Fig.\,7. In agreement with previous molecular dynamics simulation of rod-like molecular liquid crystals \cite{cifelli2006}, we find that the three sets of long-time self-diffusion coefficients exhibit a dependence on $T^{\star}$ that is well-described by an Arrhenius-like exponential law, that reads $D/D_0 \approx D^{\star} \exp(-E^{\star}/T^{\star})$, with the pre-exponential factor $D^{\star}$ and activation energy $E^{\star}$ fitting parameters. We also observe that the dependence of $D_{\perp}$ on temperature is very similar in both N and Sm phases. At a given temperature, most likely due to the packing difference between the two LC phases, the numerical value of $D_{\perp}$ is slightly larger in the N phase than in the Sm phase, but otherwise $D_{\perp}=D_{\perp}(T^{\star})$ exhibits the same exponential trend, with very similar fitting parameters, in both frames of Fig.\,7. On the other hand, the diffusion along the director is significantly slower in the Sm phase ($E^{\star} \approx 42$) than in the N phase ($E^{\star} \approx 11$) by almost one order of magnitude, most likely due to the layered structure that hampers the penetration of the rods and thus delays their diffusion along the nematic director. These findings are in qualitative agreement with former theoretical, simulation and experimental works that clarified the existence of free-energy barriers hampering the diffusion of rod-like particles through Sm layers \cite{vanroij1995, vandui1997, lettinga2007, bier2008, patti2009, patti2010}. Therefore, while SRS particles in the N phase preferentially diffuse in the direction of the nematic director, in the Sm phase they are essentially constrained in a two-dimensional space, especially at $T^{\star}<10$, where $D_{\parallel}$ is almost negligible. 

\begin{figure}[ht]
\includegraphics[width=8.5cm]{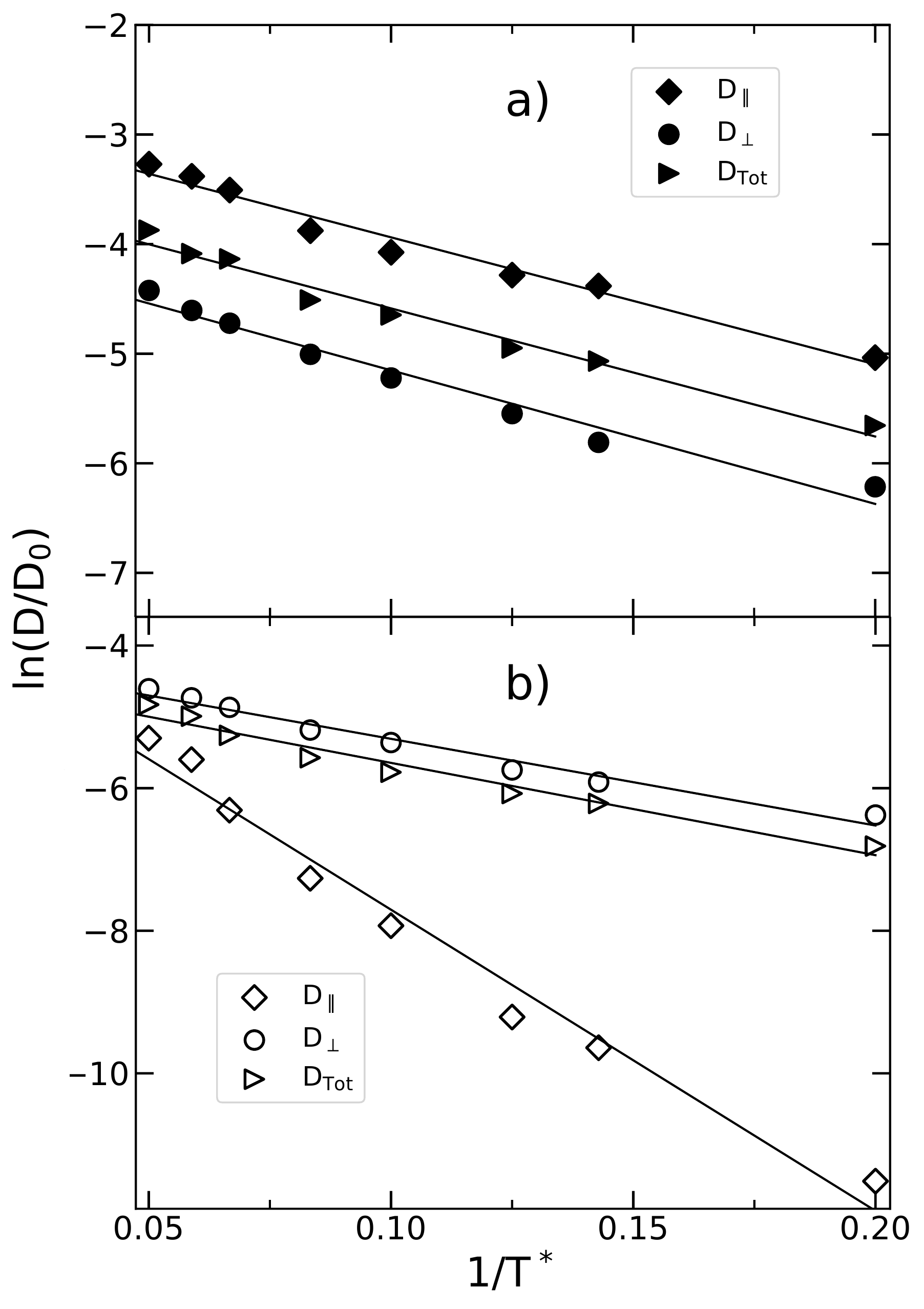}
\caption{Total ($D_{\rm Tot}$), parallel ($D_{\parallel}$) and perpendicular ($D_{\perp}$) self-diffusivities, represented with triangles, squares and circles, respectively, as a function of the reduced temperature in the N (a) and Sm (b) phases. Solid lines are exponential fits of the type $D/D_0 \approx D^{\star} \exp(-E^{\star}/T^{\star})$, with $D^{\star}=\{0.0621,\,0.0196,\,0.0328\}$ and $E^{\star}=\{11.5901,\, 12.1927,\,11.6009\}$ for parallel, perpendicular and total self-diffusivities, respectively, for the nematic states, and $D^{\star}=\{0.0308,\,0.0167,\,0.0129\}$ and $E^{\star}=\{42.243,\,12.1584,\,12.948\}$ for parallel, perpendicular and total self-diffusivities, respectively, for the smectic states.}
\end{figure}


In the light of these considerations, we now turn our attention to the probability of observing particles that displace significantly shorter or longer distances than the average particles over the same period of time. The existence of such particles, here referred to as \textit{fast} or \textit{slow}, is corroborated by the computation of the s-VHFs along the nematic director and perpendicularly to it, as given, respectively, in Eqs.\,(10) and (11). To illustrate this, we show the s-VHFs at $t/\tau = 10^4$, a time that is sufficiently long to observe the relevant dynamical features of both N and Sm phases across the whole spectrum of temperatures studied. In particular, the parallel s-VHFs of Sm LCs, shown in the top frame of Fig.\,7, display periodically peaked profiles that follow the typical layered structure of this phase. At increasing temperatures, from $T^{\star}=5$ to 20, these peaks become less and less pronounced, suggesting a more uniform  probability of finding particles at any distance along the nematic director. Nevertheless, at relatively low temperature, with the smectic layers well-defined and less prone to density fluctuations, the profiles unambiguously suggest that particles preferentially jump from layer to layer and almost no particles are observed in between. For instance, at $T^{\star}=5$, while most particles are still in their original layer (primary peak), there exist especially fast particles that succeeded in diffusing, over the same period of time, to a contiguous layer (secondary peak). This is also observed at larger temperatures, but the difference between the height of primary and secondary peaks gradually softens and eventually disappears at $T^{\star}=20$.

\begin{figure}[ht]
\includegraphics[width=8.6cm]{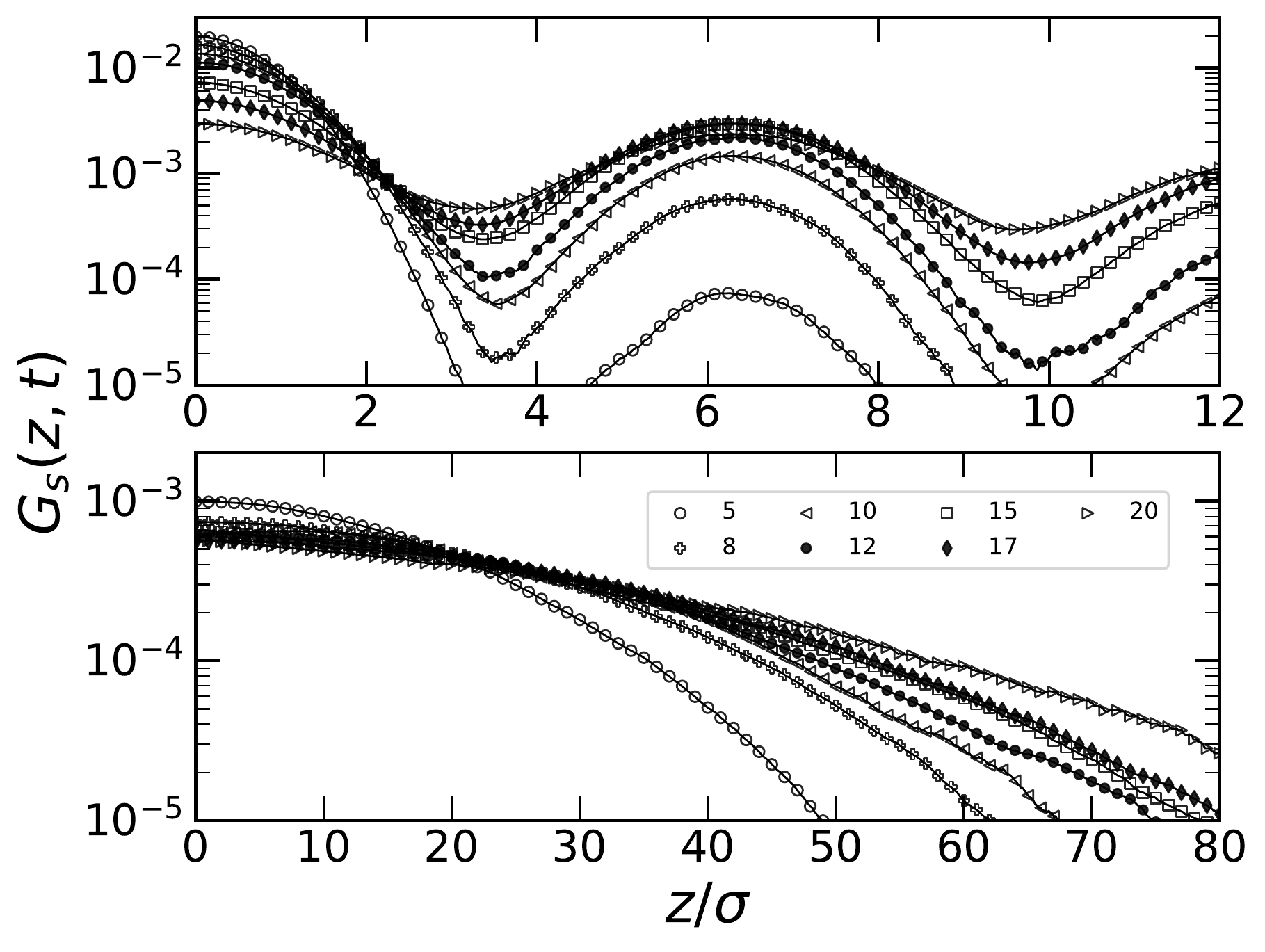}
\caption{Self-part of the van Hove function at a time $t/\tau=10^4$ in N (bottom) and Sm (top) phases along the nematic director at scaled temperatures $T^{\star}$= 5 8, 10, 12, 15, 17, and 20.}
\end{figure}

The bottom frame of Fig.\,7 reports similar s-VHFs for the N phase. In this case, profiles with a maximum at $z=0$ and monotonic decay at relatively long distances are observed. Most particles are therefore at or very close to their original position, with few of them fast enough to be displaced substantially larger distances over the same time window. With increasing temperature, more and more particles are able to move longer distances and, correspondingly, less and less are found at their original location. Finally, the perpendicular s-VHFs shown in Fig.\,8 for N (bottom frame) and Sm (top frame) phases reveal the presence of an interesting variety of particles. At $t/\tau = 10^4$, most of them have left their initial position, as indicated by the peak of the distribution. These particles coexist with others that either remained very close to their original location or displaced significantly larger distances. Upon increasing temperature, the probability of observing such slow and fast particles becomes more and more uniform and would eventually become space-independent at very large temperatures, at which the system would transform into an isotropic phase.

\begin{figure}[ht]
\includegraphics[width=8.6cm]{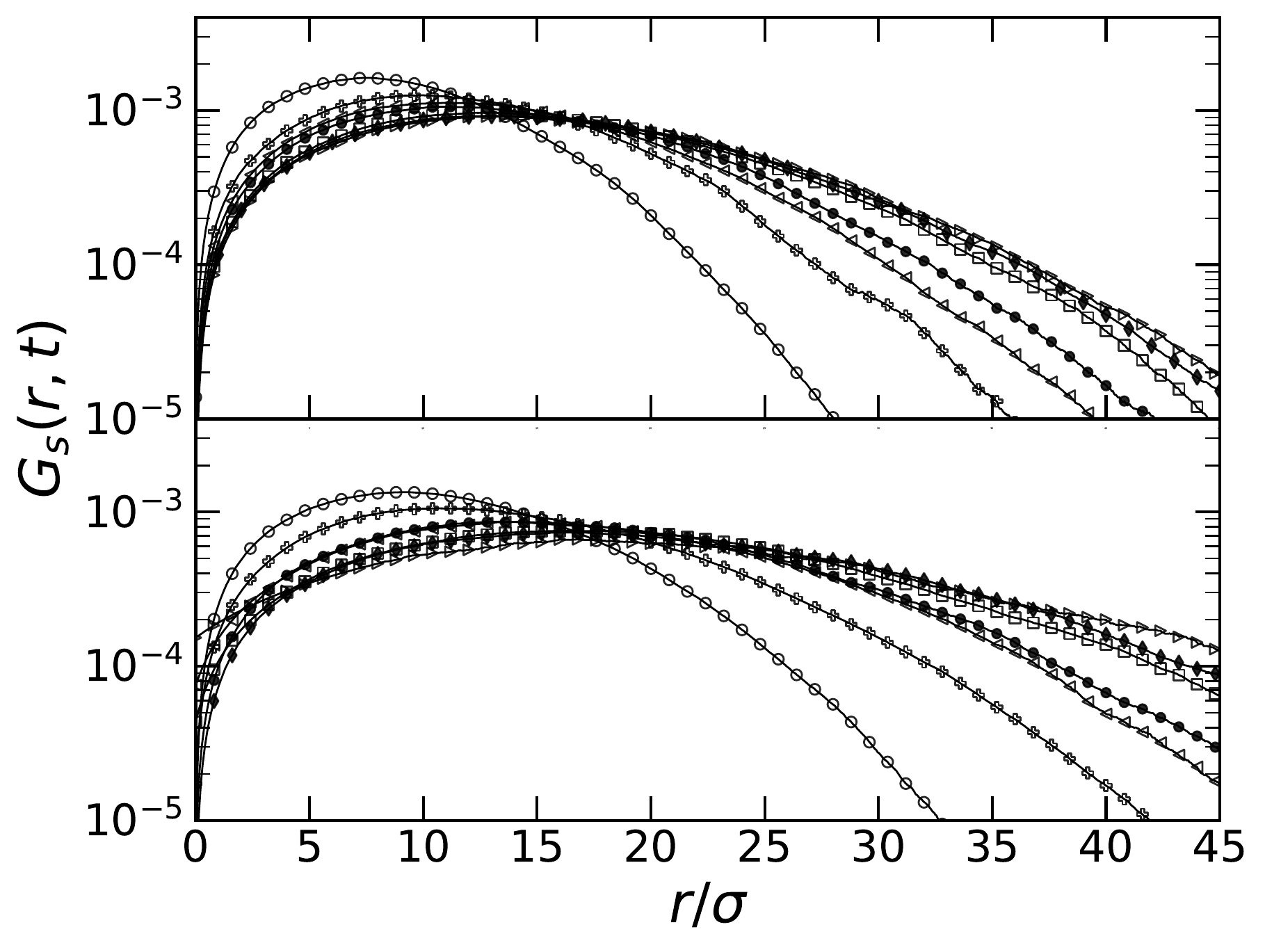}
\caption{Self-part of the van Hove function at a time $t/\tau=10^4$ in N (bottom) and Sm (top) phases in planes perpendicular to the nematic director for scaled temperatures $T^{\star}$= 5 8, 10, 12, 15, 17, and 20. Symbols as in figure 7}
\end{figure}

Temperature also plays a key role in determining the time scale of the structural relaxation of the system. This has been estimated by computing the s-ISF in direction parallel and perpendicular to $\bf{n}$. The s-ISFs of N and Sm phases are respectively shown in Figs.\,9 and 10. In both cases, they have been calculated at the wave vectors corresponding to the peak of the static structure factor, which are $\mathbf{q}=(0,0,q_z)$, with $q_z\sigma=1$ for parallel s-ISFs and $\mathbf{q}=(q_x,q_y,0)$ with $\sqrt{(q_x^2+q_y^2)}\sigma=6$ for perpendicular s-ISFs. Both LC phases exhibit a relevant difference between parallel and perpendicular relaxation, with the former taking up to 2 to 3 extra time decades. In all the cases studied, the decay of the s-ISFs closely follows a stretched-exponential function of the form $\exp[-(t/t_r)^\beta]$, typically observed in dense fluids \cite{brambilla2009}, with $t_r$ and $\beta$ fitting parameters. In particular, the exponent $\beta$ is approximately between 0.6 and 0.7 for $F_{s,xy}$, and between 0.8 and 0.9 for $F_{s,z}$, suggesting a more stretched decay in planes perpendicular to the nematic director than in the direction parallel to it. These values agree well with those reported in previous simulation of hard spherocylinders \cite{Matena2010, belli2010}. While the dependence of $\beta$ on temperature is relatively mild, the relaxation time $t_r$, defined as the time at which $F_s=1/e$, changes significantly with the temperature as can be inferred from Fig.\,11, where $\ln(t_r/\tau)$ is plotted as a function of $\ln(T^{\star})$. The so-calculated relaxation time exhibits a power-law dependence on $T^{\star}$ that holds in both Sm and N phases.

\begin{figure}[ht]
\includegraphics[width=8.8cm]{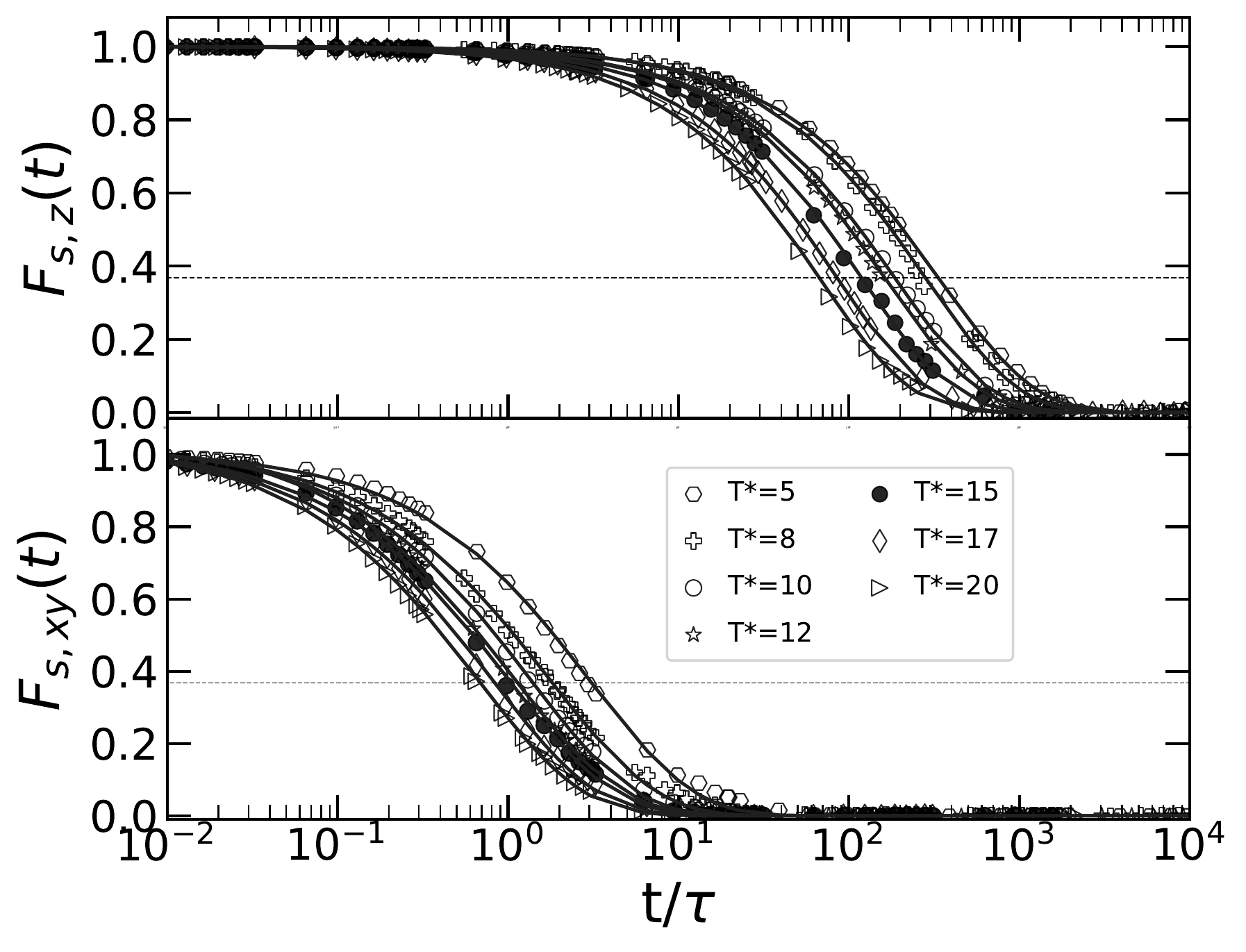}
\caption{Temperature dependence of the self-part of the intermediate scattering function in the N phase along the nematic director (top frame) and  perpendicular to it (bottom frame). Symbols indicate simulation results, while solid lines are stretched-exponential fits. All s-ISFs have been calculated at wave vectors $q_z\sigma\approx1$ and $\sqrt{q_x^2+q_y^2}\sigma\approx6$. Dashed line corresponds to the $1/e$ value where relaxation of iSF is commonly measured. }
\end{figure}

\begin{figure}[ht]
\includegraphics[width=8.7cm]{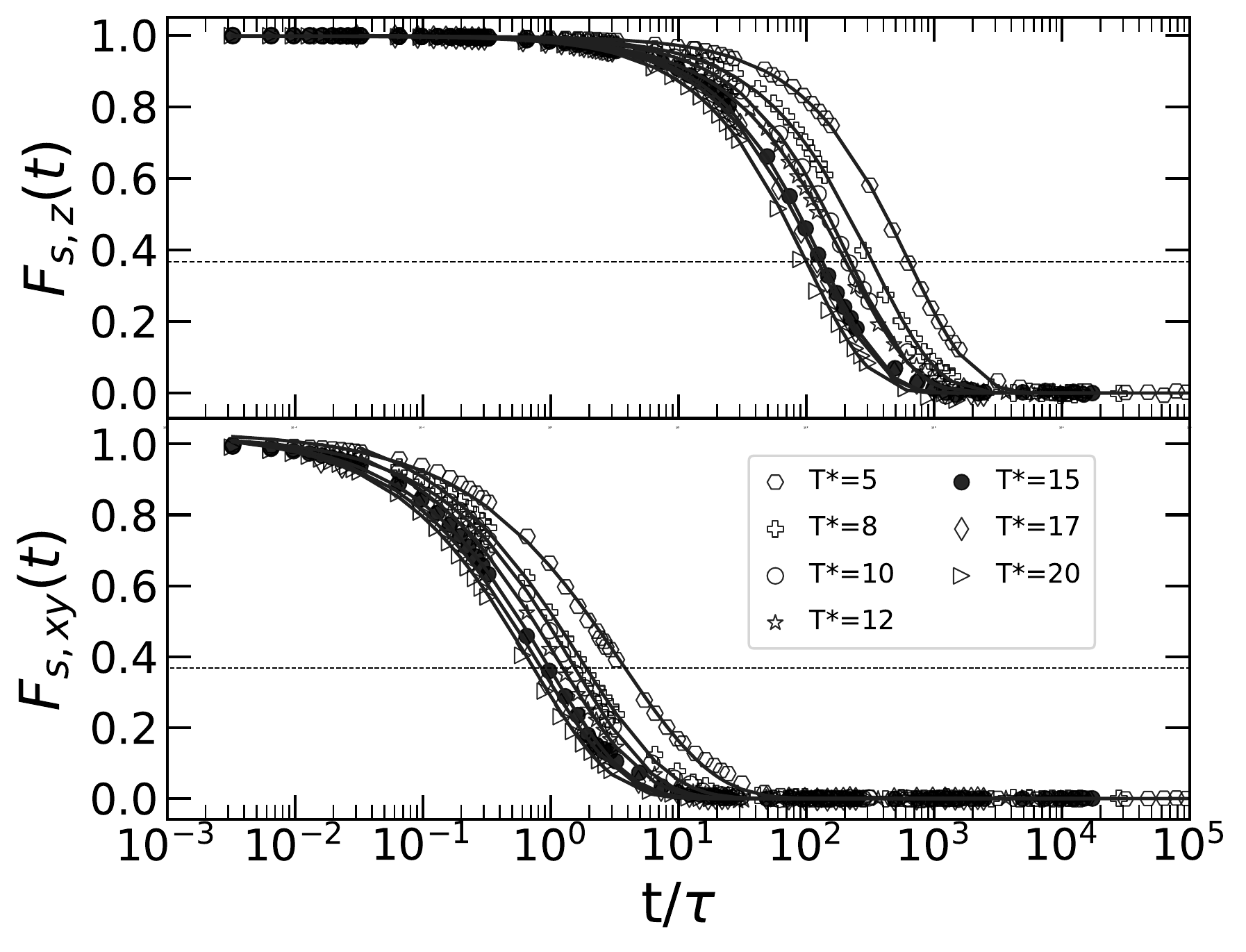}
\caption{Temperature dependence of the self-part of the intermediate scattering function in the Sm phase along the nematic director (top frame) and  perpendicular to it (bottom frame). Symbols indicate simulation results, while solid lines are stretched-exponential fits. All s-ISFs have been calculated at wave vectors $q_z\sigma\approx1$ and $\sqrt{q_x^2+q_y^2}\sigma\approx6$. Dashed line corresponds to the $1/e$ value where relaxation of iSF is commonly measured.}
\end{figure}


\begin{figure}[ht]
\includegraphics[width=8.5cm]{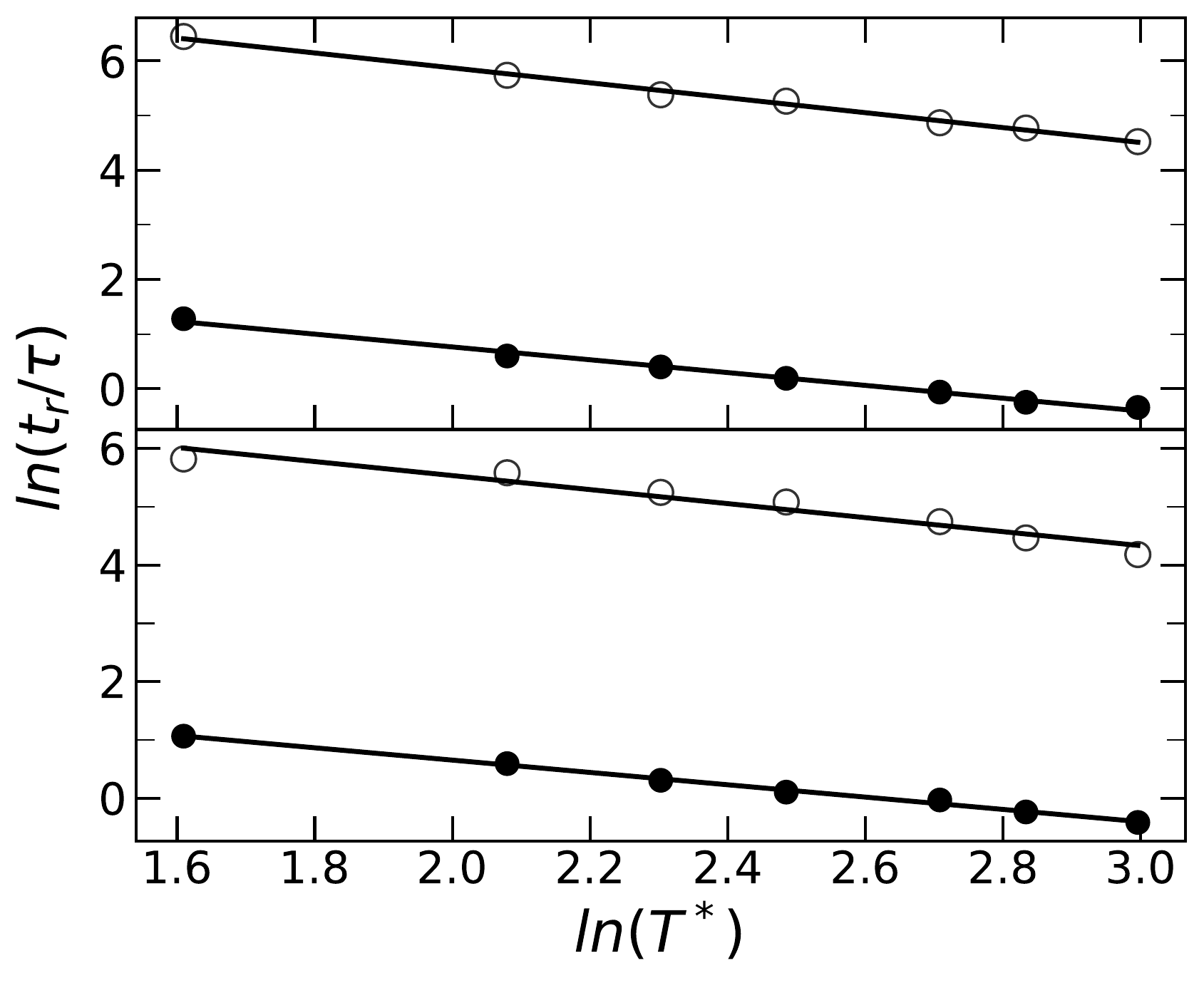}
\caption{Relaxation times for Sm (top frame) and N (bottom frame) phases. Open and solid symbols refer respectively to structural relaxation perpendicular to the nematic director and parallel to it. Solid lines are fitted functions of the form $t_r/ \tau =b\cdot {T^*}^a$ where $a$=-1.36 and $b$=2980}
\end{figure}


\section{Conclusions \protect  }
In summary, we have performed DMC simulations to investigate the dynamics of soft-repulsive rod-like particles in nematic and smectic LCs. To this end, we have calculated a spectrum of dynamical properties that helped us characterise the long-time relaxation decay in the direction of the nematic director and perpendicularly to it. In particular, the MSD was key to obtain the self-diffusion coefficients and gain an insight into their dependence on temperature, the s-VHF unveiled the existence of particles able to cover significantly longer distances than most of the particles over the same time, and the s-ISF clarified the timescales of the structural fluctuations' decay. Our results suggest a similar fluid-like diffusive behavior in the direction perpendicular to the nematic director in both N and Sm phases. By contrast, the crystal-like arrangement of the Sm phase along the nematic director determines a significantly different dynamics in this direction as compared to that observed in the N phase. In particular, rods struggle to diffuse from layer to layer as the very low value of their long-time self-diffusion coefficients reveals. The dependence of self-diffusivities on temperature follows an Arrhenius-like exponential law that suggests an especially high activation energy in the Sm phases at low temperatures. By contrast, at the same temperatures, the activation energy is up to 4 times smaller in the N phase. The dependence on temperature of the rods' dynamics in the N phase is also inferred from the inversion time, $t_i$, which is the time at which the parallel diffusion becomes faster than the perpendicular diffusion. The inversion time, which also displays an exponential dependence on temperature, is larger at low temperatures, but then gradually shorter as the system approaches the transition to the isotropic phase.

The analysis of the s-VHFs in the direction of the nematic director suggests the presence of rods that are able to displace significantly longer distances than the average particle over the same time scale. While such \textit{fast} particles are found in both the N and Sm phases, the probability distribution profiles of their parallel displacements in these phases are not the same. More specifically, periodically peaked s-VHFs are found in the Sm phases, with the peaks becoming smoother and smoother at increasing temperature. By contrast, in the N phase, no peaks are observed, but monotonically decreasing probabilities that vanish at sufficiently long distances. To some extent, the parallel s-VHFs of the N phase are very similar to those calculated perpendicularly to the nematic director, whose profiles in the N and Sm phases are almost completely indistinguishable. However, the perpendicular s-VHFs exhibit a maximum at relatively short distances that suggests the existence of especially \textit{slow} particles that, even at sufficiently long times, have displaced less than one rod diameter from their original position. The simultaneous presence of slow and fast particles contributes to determine the structural relaxation of the systems, which has been assessed by calculating the s-ISF. As observed in dense liquids, the s-ISFs exhibit a stretched exponential decay, which can take up to 3 time decades more along the nematic director than in the directions perpendicular to it.

The dynamical properties of the soft repulsive model studied here can be extended to introduce other effects that are relevant for the characterization of soft matter phases, like polydispersity in viruses \cite{Lettinga2011}  and the introduction of electrostatic interactions in models for colloids and proteins \cite{Gil2020}. Since the Mie-potential has extended the application of the Lennard-Jones model to describe properties of real substances \cite{Dufal2015}, we can expect that in similar way shifted Kihara systems of variable range, modifying the exponents 12-6, could be also relevant in order to describe liquid crystalline phases.

\section{Acknowledgments} 
D.C. acknowledges support from CONACYT for funding her PhD scholarship. A.P. acknowledges financial support from the Leverhulme Trust Research Project Grant No.\,RPG-2018-415 and the Newton Mobility Grant NMG\textbackslash R2\textbackslash 170137 awarded by The Royal Society to fund his visit at the University of Guanajuato in León.







\bibliography{main.bib}
\end{document}